\documentclass[aps,prl,superscriptaddress,twocolumn,showpacs,floatfix]{revtex4}
\usepackage{amsmath}
\usepackage{amssymb}
\usepackage{graphicx}
\usepackage{bm}
\usepackage{amsfonts}

\providecommand{\U}[1]{\protect\rule{.1in}{.1in}}
\providecommand{\U}[1]{\protect\rule{.1in}{.1in}}

\begin{document}

\title{Spin accumulation with spin-orbit interaction}
\author{Henri Saarikoski}
\email{h.m.saarikoski@tudelft.nl}
\affiliation{Kavli Institute of Nanoscience, Delft University of Technology, 2628-CJ
Delft, The Netherlands}
\affiliation{Mathematical Physics, Lund Institute of Technology, SE-22100 Lund, Sweden}
\author{Gerrit~E.~W. Bauer}
\email{g.e.w.bauer@tudelft.nl}
\affiliation{Kavli Institute of Nanoscience, Delft University of Technology, 2628-CJ
Delft, The Netherlands}
\pacs{85.75.-d, 71.10.Ca, 71.15.Mb}
\date{\today }

\begin{abstract}
Spin accumulation is a crucial but imprecise concept in spintronics. In
metal-based spintronics it is characterized in terms of semiclassical
distribution functions. In semiconductors with a strong spin-orbit coupling
the spin accumulation is interpreted as a superposition of coherent
eigenstates. Both views can be reconciled by taking into account the
electron-electron interaction: a sufficiently strong self-consistent
exchange field reduces a spin accumulation to a chemical potential
difference between the two spin bands even in the presence of spin-orbit
coupling. We demonstrate the idea on a clean two-dimensional electron gas
(2DEG) by showing how the exchange field protects a spin accumulation from
dephasing and introduces an easy-plane anisotropy.
\end{abstract}

\maketitle

Metal-based spintronics~\cite{zutic} has evolved into a mature field in
which spin phenomena are routinely exploited in versatile applications~\cite%
{awschalom}. However, integration of spin-based functionalities into
semiconductor circuits is still a pressing challenge. Much of recent
research in this area has been motivated by device concepts, such as the
seminal Datta-Das transistor \cite{dattadas}, which requires injection and
detection of spins by ferromagnetic contacts to a narrow channel of a
two-dimensional electron gas with gate-controlled spin-orbit interaction (SOI).
In spite of progress to inject, modulate, transport, and detect spin
polarization all-electrically~\cite{lou,chen,vanderwal1,folk,crooker2,Ciorga}
as well as evidence that the SOI can indeed be tuned by
external gates~\cite{nitta}, the route to a working spin transistor appears
to be still full of obstacles. In the meantime, many insights have been
obtained on the spin accumulation and its dynamics by optical methods,
especially time-dependent Kerr and Faraday rotation spectroscopy \cite%
{kikkawa2,kato,crooker,pugzlys,meier,leyland}.
Large Rashba splitting have been observed at metal surfaces by
angle-resolved photoemission \cite{lashell}, which attracts a lot of
attention recently \cite{krupin05,he}.

We define a spin accumulation as a non-equilibrium spin-polarized state
injected optically or electrically into a non-magnetic material. In
metal-based spintronics a spin accumulation is synonymous to a chemical
potential difference between spin up and down bands~\cite{spinaccu}.
However, in semiconductors the SOI prominently
affects the electronic structure and transport properties~\cite{winkler03}.
A spin accumulation is then interpreted as an intrinsically time-dependent
quantum superposition of coherent eigenstates. This difference is not just a
semantic question but essential for the functionality of spintronic devices.
Here we offer a unified mean-field theory for the spin accumulation in both
metals and semiconductors. Spin can be injected either slowly, \textit{e.g}.
by a ferromagnetic contact with small electric bias, or rapidly, \textit{e.g}%
. by pulsed optically induced excitation. We start below with a description
of spin-accumulation eigenstates that are accessible by adiabatic excitation
followed by a brief disucssion of the spin accumulation dynamics of rapidly
excited states. We illustrate the general ideas at the hand of a 2DEGs with
Rashba SOI~\cite{bychkov84}, in which the disorder-scattering lifetime
broadening is much smaller than the spin-orbit splitting at the Fermi-level.

Let us consider an infinitely extended homogeneous 2DEG. To leading order in
the electron wave vector $\mathbf{k}=-i\mathbf{\nabla}$ the Hamiltonian
including the SOI is \cite{winkler03}
$H_{0}={\hbar^{2}k^{2}}/{2m^{\ast}}+\alpha(\sigma_{x}k_{y}-\sigma _{y}k_{x}),$
where 
$m^{\ast}=m_{r}m_{e}$ is the effective electron mass, $m_{e}$ is the bare
electron mass, $\sigma_{i}$ are the Pauli matrices, and $\alpha$ is the
Rashba SOI strength parameter~\cite{bychkov84}. Electron-electron
interactions are treated within the density-matrix functional theory (DMFT)~%
\cite{zumbach} which is a generalization of the Hohenberg-Kohn-Sham
density-functional theory \cite{HKS} that can handle excited states.
Compared to the Hartree-Fock (HF) method, exchange and correlation effects
in the DFMT can be treated within local approximations. The reduced density
matrix is $\Gamma(\mathbf{z},\mathbf{z}^{\prime})=\sum_{i}^{\infty}n_{i}%
\chi_{i}\left( \mathbf{z}\right) \chi_{i}^{\dag}\left( \mathbf{z}%
^{\prime}\right)$, where $\chi_{i}$ are eigenstates of the Kohn-Sham
Hamiltonian (natural orbitals), $0<n_i<1$ are the corresponding
eigenvalues (natural occupation numbers), and $\mathbf{z}=(\mathbf{r},{\sigma}%
)$ is space-spin coordinate. We define $\Gamma_{\mathbf{s}}$ as
the subset of all density matrices which correspond to a given electron
density $\rho\left( \mathbf{r}\right)= \sum_{i{\sigma}}n_{i}\chi_{i}^{\dag}%
\left( \mathbf{z}\right) \chi_{i}\left( \mathbf{z}\right) $ and
spin-polarization $\mathbf{s}\left( \mathbf{r}\right)= \sum
_{i}n_{i}\chi_{i}^{\dag}\left( \mathbf{z}\right) {\bm\sigma}\chi_{i}\left( 
\mathbf{z}\right)/N, $ where $N$ is the total number of electrons. The
density-matrix functional is defined via minimization of the total energy $%
E[\Gamma_{\mathbf{s}}]=\min_{\Psi\lbrack\Gamma_{\mathbf{s}}]}\langle
\Psi\lbrack\Gamma_{\mathbf{s}}]|H|\Psi\lbrack\Gamma_{\mathbf{s}}]\rangle$ in
the space of all many-body wave functions that correspond to a given 
$\Gamma_{\mathbf{s}} $.

We now assume that the exact density matrix can be generated by a
non-interacting system of pseudo particles 
\begin{equation}
\lbrack H_{0}+V_{ext}+V_{H}+V_{\mathrm{xc}}]\phi_{i}=\epsilon_{i}\phi _{i},
\label{mf}
\end{equation}
where $V_{ext}$, $V_{H}$, and $V_{\mathrm{xc}}$ are the external, the
Hartree, and the exchange-correlation potential, respectively, such that $%
\Gamma_{\mathbf{s}}(\mathbf{z},\mathbf{z}^{\prime})\cong\sum_{i}^{N}f_{i}%
\phi_{i}\left( \mathbf{z}\right) \phi_{i}^{\dag}\left( \mathbf{z}^{\prime
}\right)$ with $f_{i}=\left\{ 0,1\right\}$ and 
$V_{\mathrm{xc}}(\mathbf{z},\mathbf{z}^{\prime},[\Gamma_{\mathbf{s}%
}])= \delta E_{\mathrm{xc}}[\mathbf{\Gamma_{s}}]/{\delta\Gamma _{\mathbf{s}}(%
\mathbf{z},\mathbf{z}^{\prime})}$, 
where $E_{\mathrm{xc}}$ is the exchange-correlation energy.
HF-calculations for the Rashba Hamiltonian, following Ref. \cite{chesi}%
, confirm that the effect of the SOI on the exchange
potential is negligible for small spin polarizations.
With the local approximation,
we finally arrive at $V_{\mathrm{xc}}(\mathbf{z},\mathbf{z}^{\prime
},[\Gamma _{\mathbf{s}}])\approx \delta (\mathbf{r}-\mathbf{r^{\prime }}%
)(V_{0}(\rho ,s)+J_{\mathrm{xc}}(\rho ,s)\mathbf{\hat{s}}\cdot \mathbf{%
\sigma }), $ where $\mathbf{s}=s\mathbf{\hat{s},}$ and $J_{\mathrm{xc}}(\rho
,s)$ ($\simeq J(\rho )s$ for small $s$) is the modulus of the
exchange-correlation field vector.
The scalar $V_{0}$ can be dropped in homogeneous systems.
We may approximate $J$ by the HF exchange energy of the strictly
2DEG~\cite{barth} $V_{\mathrm{xc}}\approx J_{\mathrm{x}}(\rho ,s)\mathbf{%
\hat{s}}\cdot \sigma \approx J(\rho )\mathbf{s}\cdot \sigma =-\sqrt{2}%
m_{r}/(\pi r_{s}\epsilon ^{2})\mathrm{Ry}\,\mathbf{s}\cdot \sigma ,$ where $%
r_{s}=m_{r}/(a_{B}\epsilon \sqrt{\pi \rho })$ is the dimensionless density
parameter, $\epsilon $ is the relative dielectric constant of the medium, $%
\mathrm{Ry}=13.6\;\mathrm{eV}$, and $a_{B}=0.53\;\mathrm{\mathring{A}}$.
The effective Hamiltonian is then
\begin{equation}
H\left( \mathbf{s}\right) =H_{0}+J\mathbf{s}\cdot {\bm\sigma },  \label{HJ}
\end{equation}%
where $J<0$ is the effective exchange potential, in which
correlations can be included using published parameterizations
of the correlation energy for a non-SO coupled 2DEG~\cite{LDA}.
For typical
electron densities $\rho $ the exchange energies (a few meV) are 
of the same order of magnitude as SOI energies at the Fermi-level in III--V~%
\cite{nitta} and II--VI~\cite{gui} semiconductor-based 2DEG's.

\begin{figure}[th]
\includegraphics[width=0.9\columnwidth]{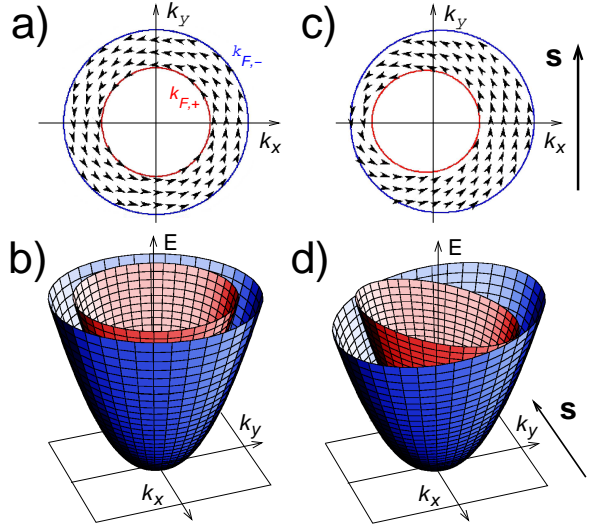}
\caption{ (Left panel) Spins (a) and
energies (b) of the ground-state spin-split bands in a non-interacting
Rashba 2DEG. The inner and outer circles correspond to the Fermi energies of
the spin-split bands.
(Right panel) Modulation of the electronic
structure in the presence of an in-plane spin
accumulation $\mathbf{s}$ by its the exchange field.
Circles in (c) are fixed-energy countours.
The shifted occupation number distributions
that minimize the energy are shown in (d).}
\label{figinplane0}
\end{figure}

\begin{figure}[th]
\includegraphics[width=0.95%
\columnwidth]{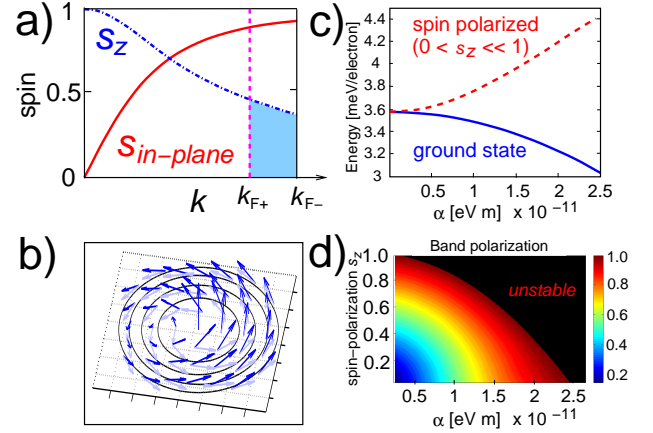}
\caption{(a-b) Spin direction of the lower spin-split ($-$) band in the
presence of a spin polarization normal to the 2DEG plane. The total spin
polarization is determined by the shaded area between $k_{F,-}$ and $k_{F,+}$%
. c) The ground state (solid line) and excited eigenstates (dashed line)
with non-zero, small spin polarization perpendicular to the 2DEG surface are
separated with an energy gap. d) Band polarization of spin-accumulation
states at fixed $J=-2\;\mathrm{meV}$. Material parameters are $\protect%
\epsilon =12.7$, $m^{\ast }=0.067m_e$, $\protect\rho =2\times 10^{15}/{%
\mathrm{m}^{2}}$.}
\label{spinstructurez}
\end{figure}

The eigenstates of the Hamiltonian in the non-polarized ground state are
split into two bands with a chiral spin pattern (Fig. \ref{figinplane0}
(a-b)).
The exchange field deforms the electronic bands and spinors as shown in Fig. %
\ref{figinplane0}(c) for in-plane and in Fig. \ref{spinstructurez}(a-b)
for perpendicular direction of an injected spin accumulation.
SO split bands of the surface states of in-plane magnetized Gd films have been
found to be deformed by the exchange potential~\cite{krupin05} similar to
Fig. \ref{figinplane0}(c).
Our task is
to find the self-consistent single-determinant eigenfunction of $H\left( 
\mathbf{s}\right) $, which according to the DMFT is unique. 
Introducing the occupation numbers $f_{\mathbf{k}\lambda }=\{0,1\}$ of the
spin-split states $\phi _{\mathbf{k}\lambda },$ 
the spin polarization reads 
\begin{equation}
\mathbf{s}=\sum_{\mathbf{k}\lambda }f_{\mathbf{k}\lambda }\langle \phi _{%
\mathbf{k}\lambda }|{\bm\sigma }|\phi _{\mathbf{k}\lambda }\rangle /N,
\label{s}
\end{equation}
where $\lambda =\{+,-\}$ is the band index. The state we are looking for
minimizes the energy under the constraint (\ref{s}), with the $f_{\mathbf{k}%
\lambda }$ as constrained variational variables. We
solve the problem either analytically in limiting cases or numerically by a
stochastic minimization method, which uses penalty functions to fix the spin
polarization and the Metropolis sampling method to find the global energy
minimum.
An unequal occupation of spin bands, $N_{-}=\sum_{\mathbf{k}%
-}f_{\mathbf{k}-}>N_{+}$ can be parameterized by a chemical potential
difference or a band polarization $p_{b}=(N_{-}-N_{+})/N>0$.
Occupations can also shift in momentum space (Fig. \ref{figinplane0}(d)).

A spin accumulation in the plane of the 2DEG can be generated
at minimized energy by shifting the Fermi circles,
which induces currents~\cite{ganichev} via a
``spin-galvanic Hall effect'': the minimum energy state at fixed $s_{x}$ is
associated with a charge current in the perpendicular $y$-direction 
\begin{equation}
j_{_{c,}y}=-e{{\frac{\alpha }{\hbar }}\left( 1+{\frac{Jm^{\ast }}{\pi \rho
\hbar ^{2}}}\right) }s_{x}.
\end{equation}%
Since electrons move with constant drift velocity, there are no intrinsic
spin-Hall currents~\cite{sinova}. Non-equilibrium spin currents are induced,
but vanish to first order in $s_{x}$.

A spin accumulation perpendicular to the 2DEG surface can
be generated by the exchange field that pops the in-plane spin textures out
of the plane (Fig.~\ref{spinstructurez}(b)). However, the SOI
counteracts the spin-alignment and the band polarization $p_{b}$
must be increased from that of the ground state to support a spin
accumulation~\cite{footnote}. Consequently, excited eigenstates
corresponding to a finite $s_{z}$ are, in contrast to the in-plane case,
separated from the non-polarized ground state by a finite energy gap 
\begin{equation}
E_{g}=\left( \frac{\alpha m^{\ast }}{\hbar J}\right) ^{2}\left( J+\frac{%
\hbar ^{2}\pi \rho }{m^{\ast }}\right) +{\mathcal{O}}(\alpha ^{4}).
\label{gapenergy}
\end{equation}%
The divergence in $E_{g}$ at $J\rightarrow 0$ reflects the absence of an $%
s_{z}$ component in the non-interacting Rashba model. The gap is shown in
Fig. \ref{spinstructurez}(c) in which Eq. (\ref{gapenergy}) corresponds to
the low $\alpha $ limit of the energy difference. This gap energy must be
overcome to achieve spin-polarized eigenstates at arbitrarily small $%
s_{z}\neq 0$. Except for this singular behavior at $s_{z}=0$ we find that
the energy of eigenstates is isotropic (to second order in $s$) to the
numerical accuracy for $s_{z}\neq 0$. We suspect that there may be a more
general physical reason behind this out-of-plane isotropy. The contribution
of spin-galvanic currents to $E({\mathbf{s}})$ is not significant for
material parameters shown in Fig. \ref{spinstructurez}. 

The maximum spin accumulation that can be accommodated perpendicular to the
2DEG surface is determined by the total spin polarization of a single
occupied band $\left( p_{b}=1\right) $. In the exchange-only approximation
the self-consistency criterion (\ref{s}) can be fulfilled only when 
\begin{equation}
\alpha<2|J|/k_{F}=2|J|/\sqrt{4\pi\rho}=0.32\;\mathrm{eV}\;\mathrm{nm }%
/\epsilon,  \label{stabilitylimit}
\end{equation}
Figure \ref{spinstructurez}(d) shows the stability limit as a function of $%
s_z$.

Per definition, eigenstates do not dephase. The dynamics of the
semiclassical spin accumulations discussed above is therefore governed by
the Bloch equation $\mathbf{\dot{s}}=-\gamma \mathbf{s\times B}_{\mathrm{eff}%
}\mathbf{-s/}T_{1}$ \cite{zutic}, in which $\mathbf{B}_{\mathrm{eff}%
}=-\partial E\left( \mathbf{s}\right) /\partial \mathbf{s}$ and $T_{1}$ is
the spin relaxation time.  Due to the singular anisotropy of 
$E\left( \mathbf{s}\right) $ the Bloch equation is mathematically not well
defined in the mean-field theory employed here.
Qualitatively, the
absence of an angle dependence of $E({\mathbf{s}})$ (for $s_z \neq 0$)
implies that exchange-stabilized spin accumulations do not feel an internal
SOI field and precess only when an external magnetic field is applied. A
spin accumulation exactly in the 2DEG plane is
trapped and precesses around an in-plane external magnetic field that
exceeds a threshold value governed by the energy gap (\ref{gapenergy}).

A large phase space available for scattering processes makes a large spin
accumulation susceptible to fast decay. Therefore the stability limit of
eigenstates (\ref{stabilitylimit}) is not a sharp phase boundary. For not
too highly excited systems, the Dyakonov-Perel \cite{dyakonov} mechanism by
random scattering at defects is believed to be the dominant source of finite 
$T_{1}$ spin life times in clean systems. Since in an exchange stabilized
2DEG the precession in the SO field is suppressed, the efficiency of the
Dyakonov-Perel mechanism is strongly diminished for systems in the clean
limit considered here. The opposite (dirty) regime can be handled by
spin-coherent kinetic \cite{glazov,wengwu} and diffusion~\cite%
{burkov,mishchenko,inanc} equations or numerical simulations \cite%
{leyland,vanderwal2}.

The Datta-Das transistor is a spin valve consisting of a 2DEG spacer with
transparent ferromagnetic contacts~\cite{dattadas}.
Even when the magnetizations of
the two electrodes are parallel to each other, transport depends on the
magnetization direction when exchange is taken into account. When
magnetizations are oriented in the current direction, the spin accumulation
can be injected into the 2DEG as an eigenstate with shifted distributions
(Fig. \ref{figinplane0}(c)), which does not precess
and, hence, does not react to a gate voltage that modulates the
SOI. For magnetizations not in the 2DEG plane spin accumulation eigenstates
are separated from the ground state by an energy gap and spin cannot be
injected adiabatically at low energies. A non-adiabatic spin injection, on
the other hand, could lead to a coherent superposition of eigenstates and
spins would precess in the SOI field, as envisioned by Datta and
Das. 

\begin{figure}[tbp]
\includegraphics[width=0.75\columnwidth]{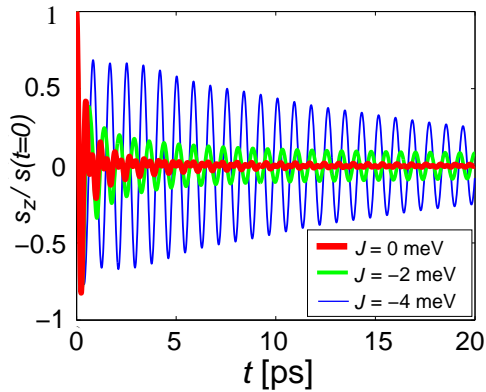}
\caption{Oscillation and dephasing of a spin ensemble in a SO-field. The
state is a coherent superposition of spin eigenstates, $s(t=0)=10\%$, and $%
\protect\alpha =4\cdot 10^{-11}\;\mathrm{eV}\;\mathrm{m}$. }
\label{fig:precess}
\end{figure}

In optical pump and probe experiments, spin-polarized electrons and holes
are generated by short resonant pulses of circularly polarized light,
followed by fast thermalization and spin relaxation of the holes.
A fast excitation creates a coherent superposition of individual
spin eigenstates which dephases with time. 
We calculate the dynamics of the spin accumulation from the initial state $%
\psi _{0}$ for times $t>0$ from $|\psi (t)\rangle =\int dt\,e^{{i}%
H(t)t/\hbar }|\psi _{0}\rangle ,$ where the Hamiltonian depends on $t$ by
the exchange field $J\mathbf{s}\left( t\right) $. The state at $t$ is solved
iteratively 
\begin{equation}
|\psi (t)\rangle \approx e^{{i}H(t-\Delta t)\Delta t/\hbar }\ldots e^{{i}%
H(t=\Delta t)\Delta t/\hbar }e^{{i}H(t=0)\Delta t/\hbar }|\psi _{0}\rangle
\end{equation}%
for short time steps $\Delta t \simeq \mathrm{1\; fs}$.
We assume that dephasing is fast compared to the spin relaxation processes
so that the occupation numbers are unchanged. Figure \ref{fig:precess} shows
the time evolution of the spin accumulation, which has been excited at time $%
t=0$ 
into a coherent superposition of eigenstates. A single spin oscillates in the
SO-field by a frequency $\omega _{\mathrm{SO}}=2\alpha k/\hbar .$ The spin
polarization excited over a finite band width is therefore expected to decay
on the scale of the dephasing time $T_{2}$, that decreases with increasing $%
s.$ However, a strong exchange field aligns spins along a common axis and
synchronizes spin precession which protects the spin polarization from
dephasing. The exchange-induced enhancement of $T_{2}$ becomes significant
when the exchange-splitting, which is proportional to $s,$ becomes of the
same order of magnitude as the spin-orbit splitting. Such an effect has been
observed in experiments \cite{stich}, but can be explained by the
exchange effect in the dirty limit as well \cite{glazov,wengwu}. 

In the space of the parameters provided by material and excitation conditions
the spin accumulation features both semiclassical and quantum properties.
The exchange field and thus the spin accumulation can be engineered by
electron density, excitation intensity, spin direction and electric
currents, and should therefore be considered in advanced spintronic device
concepts~\cite{lou,meier}. Our theoretical framework is general and can be
extended to treat three-dimensional, inhomogeneous, and finite systems as
well as the Dresselhaus SOI~\cite{dresselhaus55}.
The electronic structures of other non-magnetic
conductors with significant SOI, \textit{e.g}. hole gases in doped
semiconductors or non-magnetic transition metals, are more complicated,
but still amenable to a computational implementation of our method.

This work has been supported by Stichting FOM and NWO.
H.S. acknowledges support from the Academy of Finland. We acknowledge useful
discussions with \.{I}nan\c{c} Adagideli and Klaus Capelle.


\begin{thebibliography}{99}
\bibitem{zutic} I. \v{Z}uti\'{c}, J. Fabian, and S. Das Sarma, Rev. Mod.
Phys. \textbf{76}, 323 (2004).

\bibitem{awschalom} D. D. Awschalom and M. E. Flatt\'{e}, Nature Phys. 
\textbf{3}(3), 153 (2007).

\bibitem{dattadas} S. Datta and B. Das, Appl. Phys. Lett. \textbf{56}, 665
(1990).

\bibitem{lou} X. Lou, C. Adelmann, S. A. Crooker, E. S. Garlid, J. Zhang, K.
S. Madhukar Reddy, S. D. Flexner, C. J. Palmstr\o m, and P. A. Crowell,
Nature Phys. \textbf{3}, 197 (2007).

\bibitem{chen} P. Chen, J. Moser, P. Kotissek, J. Sadowski, M. Zenger, D.
Weiss, and W. Wegscheider, Phys. Rev. B \textbf{74}, 241302(R) (2006).

\bibitem{vanderwal1} E. J. Koop, B. J. van Wees, D. Reuter, A. D. Wieck, and
C. H. van der Wal, arXiv:0801.2699.

\bibitem{folk} S. M. Frolov, A. Venkatesan, W. Yu, S. Luescher, W.
Wegscheider, and J. A. Folk, arXiv:0801.4021.

\bibitem{crooker2} S. A. Crooker, M. Furis, X. Lou, C. Adelmann, D. L.
Smith, C. J. Palmstr\o m, and P. A. Crowell, Science \textbf{309}, 2191
(2005).

\bibitem{Ciorga} M. Ciorga, A. Einwanger, U. Wurstbauer, D. Schuh, W.
Wegscheider, and D. Weiss, arXiv:0809.1736

\bibitem{nitta} J. Nitta, T. Akazaki, H. Takayanagi, and T. Enoki, Phys.
Rev. Lett. \textbf{78}, 1335 (1997).

\bibitem{kikkawa2} J. M. Kikkawa, I. P. Smorchkova, N. Samarth, and D. D.
Awschalom, Science \textbf{277}, 1284 (1997).

\bibitem{crooker} S. A. Crooker, D. D. Awschalom, and N. Samarth, IEEE J.
Sel. Top. Quantum Electron. \textbf{1}, 1082 (1995).

\bibitem{pugzlys} A. Pug\v{z}lys, P. J. Rizo, K. Ivanin, A. Slachter, D.
Reuter, A. D. Wieck, C. H. van der Wal, and P. H. M. Loosdrecht, J. Phys.:
Condens. Matter \textbf{19}, 295206 (2007).

\bibitem{leyland} W. J. H. Leyland, R. T. Harley, M. Henini, A.J.
Shields, I. Farrer, and D. A. Ritchie, Phys. Rev. B \textbf{77}, 205321
(2008).

\bibitem{meier} L. Meier, G. Salis, I. Shorubalko, E. Gini, S. Sch\"{o}n,
and K. Ensslin, Nature Phys. \textbf{3}, 651 (2007).

\bibitem{kato} Y. K. Kato, R. C. Myers, A. C. Gossard, and D. D. Awschalom,
Science \textbf{306}, 1910 (2004).

\bibitem{lashell} S. LaShell, B. McDougall, and E. Jensen, Phys. Rev. Lett.
{\bf 77}, 3419 (1996).

\bibitem{krupin05} O. Krupin {\em et al.}, Phys. Rev. B {\bf 71},
201403(R) (2005).

\bibitem{he}
K. He, T. Hirahara, T. Okuda, S. Hasegawa, A. Kakizaki, and Iwao Matsuda
Phys. Rev. Lett. {\bf 101}, 107604 (2008); J. Hugo Dil, F. Meier,
J. Lobo-Checa, L. Patthey, G. Bihlmayer, and J.
Osterwalder, Phys. Rev. Lett. {\bf 101}, 266802 (2008);
E. Frantzeskakis, S. Pons, H. Mirhosseini, J. Henk, C. R. Ast, and M. Grioni,
Phys. Rev. Lett. {\bf 101}, 196805 (2008).

\bibitem{spinaccu} A. G. Aronov, Pis'ma Zh. Eksp. Teor. Fiz. \textbf{24}, 37
(1976) [JETP Lett. \textbf{24}, 32 (1976)]; M. Johnson and R. H. Silsbee,
Phys. Rev. Lett. \textbf{55}, 1790 (1985); F. J. Jedema, A. T. Filip, and B.
J. van Wees, Nature \textbf{410}, 345 (2001).

\bibitem{winkler03} Roland Winkler, Spin-orbit coupling effects in
two-dimensional electron and hole systems, Springer Tracts in Modern Physics 
\textbf{191}, Springer-Verlag (2003).

\bibitem{bychkov84} Y.~A. Bychkov and E.~I. Rashba, J. Phys. C: Solid State
Phys. \textbf{17}, 6039 (1984) [Y.~A. Bychkov, E.~I. Rashba, JETP Lett. 
\textbf{39}(2), 78 (1984)].

\bibitem{zumbach} G. Zumbach and K. Maschke, J. Chem. Phys. \textbf{82},
5604 (1985).

\bibitem{HKS} W. Kohn and L. J. Sham, Phys. Rev. \textbf{140}, A1133 (1965).

\bibitem{barth} U. von Barth and L. Hedin, J. Phys. C: Solid State Phys. 
\textbf{5}, 1629 (1972).

\bibitem{chesi} S. Chesi, G. Simion, and G. F. Giuliani, cond-mat/0702060
(2007).

\bibitem{LDA} C. Attaccalite, S. Moroni, P. Gori-Giorgi, and G. B. Bachelet,
Phys. Rev. Lett. \textbf{88} 256601 (2002).

\bibitem{gui} Y. S. Gui, J. Liu, V. Daumer, C. R. Becker, H. Buhmann, and L.
W. Molenkamp, Physica E \textbf{12}, 416 (2002).


\bibitem{ganichev} S. D. Ganichev {\em et al.}, Nature \textbf{417}, 153-156 (2002).

\bibitem{sinova} J. Wunderlich, B. Kaestner, J. Sinova, and T. Jungwirth,
Phys. Rev. Lett. \textbf{94}, 047204 (2005).


\bibitem{footnote} We find that at large SOI it becomes energetically
favorable to instead depopulate a large fraction of states in the majority
spin band around $k=0$.

\bibitem{dyakonov} M. I. Dyakonov and V. I. Perel, Fiz. Tverd. Tela \textbf{%
13}, 3581 (1971) [Sov. Phys. - Solid State \textbf{13}, 3023 (1971)].

\bibitem{glazov} M. M. Glazov and E. L. Ivchenko, JETP Lett. \textbf{75},
403 (2002)

\bibitem{wengwu} M. Q. Weng and M. W. Wu, Phys. Rev. B \textbf{68}, 075312
(2003).

\bibitem{burkov} A. A. Burkov, A. S. N\'{u}n\~{e}z, and A. H. MacDonald,
Phys. Rev. B \textbf{70}, 155308 (2004).

\bibitem{mishchenko} E. G. Mishchenko, A. V. Shytov, and B. I. Halperin,
Phys. Rev. Lett. \textbf{93}, 226602 (2004).

\bibitem{inanc} \.{I}. Adagideli and G. E. W. Bauer, Phys. Rev. Lett. 
\textbf{95}, 256602 (2005).

\bibitem{vanderwal2} E. J. Koop, B. J. van Wees, and C. H. van der Wal,
arXiv:0804.2968.

\bibitem{stich} D. Stich {\em et al.}, Phys. Rev. Lett. \textbf{98},
176401 (2007).

\bibitem{dresselhaus55} G. Dresselhaus, Phys. Rev. \textbf{100}, 580 (1955).

\end{thebibliography}
\end{document}